\documentclass[10pt]{article}
\usepackage{amssymb}
\usepackage{amsmath}
\usepackage{graphicx}

\newcommand{\ga}{\alpha}

\newcommand{\gd}{\delta}
\newcommand{\gl}{\lambda}
\newcommand{\gs}{\sigma}

\newcommand{\tu}{{\bar u}}
\newcommand{\tv}{{\bar v}}
\newcommand{\trho}{{\bar \rho}}
\newcommand{\tx}{{\bar x}}

\newcommand{\rf}[1]{(\ref{#1})}
\begin{document}
%%%%%%%%%%%%%%%%%%%%%%%%%%%%%%%%%%%%%%%%%%%%%%%%%%%%%%%%%%%%%%%
\begin{flushright}
hep-th/0409131\\
ITFA-2004-39
\end{flushright}

 \centerline{\Large\bf Black holes from colliding
wavepackets}
\bigskip
\centerline{\large Steven B.~Giddings\footnote{\tt giddings@physics.ucsb.edu}}
\medskip
\centerline{\em Department of Physics\footnote{Permanent address}
and Kavli Institute for Theoretical Physics} \centerline{\em
University of California, Santa Barbara, CA 93106}
\bigskip\bigskip
\centerline{\large Vyacheslav S.~Rychkov\footnote{\tt
rychkov@science.uva.nl}}
\medskip \centerline{\em Institute for
Theoretical Physics, University of Amsterdam} \centerline{\em
1018XE Amsterdam, The Netherlands}

\bigskip\bigskip
\begin{abstract}
\noindent
Arguments for black hole formation in collisions of
high-energy particles have rested on the emergence of a closed
trapped surface in the classical geometry of two colliding
Aichelburg-Sexl solutions.  Recent analysis has, however, shown
that curvatures and quantum fluctuations are large on this
apparent horizon, potentially invalidating a semiclassical
analysis.  We show that this problem is an artifact of the
unphysical classical point-particle limit:  for a particle
described by a quantum wavepacket, or for a continuous matter
distribution, trapped surfaces indeed form in a controlled regime.
\end{abstract}

\date{September 2004}
%%%%%%%%%%%%%%%%%%%%%%%%%%%%%%%%%%%%%%%%%%%%%%%%%%%%

\section{Introduction}
Production of black holes in high-energy collisions has long been a topic of interest\cite{Penrose,D'Eath}.
With the discovery of large extra dimension\cite{ADD,AADD} or warped compactification\cite{RS,GKP} scenarios
that lower the fundamental Planck scale to the TeV scale, it became clear that black holes might be
experimentally accessible in accelerators; for early discussion of this possibility see \cite{BF,Katz}.
Even more astounding,
production rate estimates \cite{GT,DL} showed that in the most optimistic version of these scenarios
black holes could be produced copiously at LHC, at rates up to about 1 BH/s.
These predictions were based on the so called geometric
cross section estimate:
\begin{equation}\label{geom}
    \sigma(2\to BH) \sim R_h^2\sim \frac
1 {M_p^2} \left[ \frac{E}{ M_p}\right ]^{\frac 2{D-3}}\ .
\end{equation}
Here $D$ is the total number of  dimensions, $M_p\sim 1$ TeV
is the fundamental $D$-dimensional Planck scale, $E$ is the C.M. collision energy, which also provides the approximate black hole mass, and
$R_h$ is the horizon radius of the
$D$-dimensional Schwarzschild BH of mass $E$.

The estimate \rf{geom} is based on the simple classical gravity
intuition, encoded for example in the ``hoop conjecture
\cite{Thorne}," that gravitational collapse and BH formation will
occur if the colliding particles pass within distance $\lesssim
R_h$ from each other. This intuitive argument was made precise in
\cite{EG} where it was argued that for impact parameters $\lesssim
R_h$ a closed trapped surface (CTS) forms in the collision
spacetime; for $D>4$ numerical verification of this argument  was
supplied in \cite{YN}. Black hole formation then follows from this
as a consequence of the singularity theorems and cosmic censorship
conjecture of classical gravity.

Recently \cite{bh1,R}, one of us has found a  loophole in this
classical argument for black hole formation. Specifically, the
analysis of \cite{EG} investigated the collision of two
Aichelburg-Sexl\cite{AS} solutions, corresponding to the
gravitational shock-waves of ultra-relativistic classical point
particles. However, as \cite{bh1} points out, and as we will
review, in the intersection of the planes of the two shock-waves,
there is a divergent curvature invariant.\footnote{Concerns about
high curvature were also previously expressed in \cite{Hsu}.} This
undermines the classical analysis, as do related arguments
\cite{R} about large fluctuations of the gravitational field.

This paper will address these issues, taking into account features that
should be present in a more complete semiclassical analysis: specifically,
the finite width of any wavepacket describing collision of quantum particles.
As a consequence, we will argue that the geometric value of the black hole p
roduction cross section remains robust.

\section{Classical gravity description}
\label{2}
 Let us briefly review the classical gravity description
of the BH production process as presented in \cite{EG}.

A classical BH may form in a collision of two particles with total
C.M.\ energy $E\gg 1$. (Here and below the $D$-dimensional Planck
units are used,  with $8\pi G=1$.) We will assume that the size of
the created BH is much smaller than the size of the large extra
dimensions (this assumption is satisfied in typical TeV-scale
gravity scenarios). We will also neglect the brane tension. Under
these assumptions, the BH production can be considered as
happening in flat $D$-dimensional space.

BH production was described in \cite{EG} by considering two
ultrarelativistic point particles in a grazing collision with an
impact parameter $b$. The gravitational field of one such particle i
s given by the Aichelburg-Sexl metric\cite{Pirani,AS,D'Eath,DH},
\begin{equation}
ds^2 = -d\tu d\tv + d\tx^{i2} + \Phi(\tx^i) \delta(\tu) d\tu^2\ .\label{AiSe}
\end{equation}
Here ${\bar u}= t-z$, ${\bar v}=t+z$, and  $\Phi$ depends only on the radius
in the $D-2$ transverse directions, $\tx^i$, ${\trho}=
\sqrt{\tx^i\tx_i}$,
and takes
the form
\begin{eqnarray}
\Phi&=& -\frac E\pi \log(\trho)\ ,\ D=4\ ;\label{phifour}\\
\Phi&=& \frac{2E}{(D-4)\Omega_{D-3}\trho^{D-4}}\ ,\
D>4,\label{phidef}
\end{eqnarray}
where $\Omega_{D-3}$ is the volume of the unit $(D-3)$-sphere.

This solution has
curvature concentrated on the plane transverse to the direction of
motion.  Indeed, the only nonzero components of
the Riemann tensor for the right-moving particle are \cite{bh1}
\begin{equation}
\label{Riemann} R_{uiuj} =  -\frac 12\,\delta(\bar
u)\,\frac{\partial^2}{\partial \tx_i\partial \tx_j}\Phi\ .
\end{equation}

This field should be superposed with the similar field of the
left-moving particle, shifted by $b$ in the transverse direction.
The resulting field is valid outside the region $\tu,\tv>0$, where the
colliding shocks start influencing each other. The metric in this
region should in principle be found by solving Einstein's
equation, but it remains unknown even in the simplest $b=0$ case.
Thus, BH formation may be concluded only indirectly, looking for a
CTS in the known part of the spacetime.  The CTS constructed  in
\cite{EG} (with numerical solution provided for $D>4$ in \cite{YN}) lies in the
union of pre-collision parts of the shock planes $\tu=0$ and $\tv=0$.
It looks like two roughly elliptically shaped surfaces, glued together
across the shock waves at the collision plane $\tu=\tv=0$.
The size of the surface is comparable to the Schwarzschild radius,
\begin{equation}\label{}
 R_h\sim E^{\frac 1{D-3}}\ .
\end{equation}
The maximal value of the impact parameter $b_{\max}\sim R_h$ for
which a CTS is found to exist leads to the geometric cross section
estimate \rf{geom}.

The issue noticed in \cite{bh1} follows directly from the curvature \rf{Riemann}.  For a single such shock wave,
there is no divergent curvature invariant except at the precise location of the particle.
But for the full solution corresponding to the combined shock waves, there is a divergent curvature
invariant in the intersection of the two shocks at $\tu=\tv=0$.
Indeed, combining \rf{Riemann} with the oppositely-moving counterpart gives
\begin{equation}
\label{Rsq} (R_{\mu\nu\gl\gs})^2 \sim
({E^2}/{\trho^{2D-4}})\,\gd(\tu)\,\gd(\tv)
\end{equation}
yielding a divergent result throughout the intersection $\tu=\tv=0$.
Since classical gravity must fail at such a curvature singularity, and since
the closed trapped surface passes through the offending region,
the argument for black hole formation, based on classical evolution
of the trapped surface, is not on solid ground.

\section{From classical to semiclassical}

Clearly the collision of two photons on Earth does not produce a
gravitational singularity in the next galaxy.  The effects that
remove the singularity in \rf{Rsq} come from the fact that
particles are intrinsically quantum, and thus have a
quantum-mechanical width.  We will argue that taking this into
account suffices to reinstate the robustness of the argument for
black hole formation.

\subsection{Wavepackets}

To go beyond the point-particle approximation, we must take into
account the limitations imposed by quantum theory.  In particular,
for a highly relativistic particle with $p_z\sim E$, there is a
typical position uncertainty  $\Delta z\gtrsim 1/E$.

Of course, the position uncertainty may be even greater. For
example, the fundamental quantum limits may be accounted for by
considering minimal-uncertainty wavepackets of the form
\begin{equation}\label{minunc}
\psi(z,t) \sim \exp\left\{ -\frac{[z-z(t)]^2}{\Delta z^2}
-ip_z[z-z(t)]\right\}
\end{equation}
(with appropriate generalization for transverse coordinates). If
we consider the collision of two such wavepackets, with widths
$\Delta z$, the condition for us to be able to still use the above
geometrical reasoning is that the wavepacket width be much less
than the Schwarzschild radius,
\begin{equation}
\Delta z\ll R_h\ .
\end{equation}
This is true because at distances large as compared to $\Delta z$
the resulting solution will be a small perturbation of the
point-particle spacetime.  In particular, the $\gd$-function in
\rf{Riemann} will be smeared over an interval of length $\sim
\Delta z$ (so that the shock acquires finite width), and the
collision spacetime will still contain a CTS. Strictly speaking
this relies on the argument, given in \cite{EG}, that the CTS can
be deformed out of the shock planes.

Note also that in a given experiment, for example at LHC, one may
also be in practice working with wavepackets of a large, even
macroscopic size.  In this case we can think of decomposing these
wavepackets into smaller wavepackets of size $\Delta z$ satisfying
the previous two limits:
\begin{equation}\label{limitss}
    E^{-1}\lesssim \Delta z \ll R_h
\end{equation}
This subdivision can be carried out in such a manner that
different small wavepackets correspond to almost orthogonal
states.  Combining the contributions of the smaller wavepackets
still results in the geometric cross section.

\subsection{Curvature}

In the context of collisions of such wavepackets, we can revisit
the question of large curvature:  namely, can we choose a
wavepacket size such that
\begin{equation}\label{limits}
   E^{-1}\lesssim \Delta z \ll R_h
\end{equation}
that avoids the large curvature of \rf{Rsq}?

To answer this, note that for wavepackets of the form \rf{minunc},
the $\delta$-function in \rf{Riemann} gets replaced by a Gaussian,
with maximum strength ~$1/\Delta z$.  This follows for example
from the fact that the integral of the curvature across the shock
should be independent of the width of the wavepacket.  Thus
\rf{Rsq} gets replaced by
\begin{equation}
 (R_{\mu\nu\gl\gs})^2 \lesssim ({E^2}/{\trho^{2D-4}})\,\Delta z^{-2}\ .
 \end{equation}
Evaluating this in the vicinity of the trapped surface, $\trho\sim R_h$, gives
\begin{equation}
(R_{\mu\nu\gl\gs})^2 \lesssim R_h^{-2}\Delta z^{-2}\ .
\end{equation}
{}From this we see that the curvature can be kept small for
\begin{equation}
\label{lim1}
\Delta z\gg 1/R_h\ ,
\end{equation}
which is compatible with the allowed range \rf{limits}.
Thus it is always possible to choose wavepackets small as compared to the size of the closed trapped surface, and such that the curvature remains small in the vicinity of the closed trapped surface.
This avoids the singular curvatures found in \cite{bh1,R}.

\subsection{Quantum fluctuations}

Another test of the semiclassical description of scattering is to estimate the strength of quantum fluctuations in the gravitational field relative to the semiclassical solution.
For the
validity of the semiclassical description these fluctuations
should be small compared to the field itself. This is also
equivalent to requiring that the occupation numbers of the
background graviton field be large---the condition proposed in \cite{R}.

We will focus on the part of the spacetime near the shock front
$\tu=0$, since everywhere else spacetime is flat. The form of the shock-wave metric given in \rf{AiSe}
is inconvenient
to use, because of its divergent components. We will instead use
the metric
\begin{equation}
ds^2=-du\, dv+\left[1+\frac{(D-3)E}{\Omega_{D-3}\, \rho^{D-2}}\,
u\theta(u)\right]^2d\rho^2 +\left[1-\frac{E}{\Omega_{D-3}\,
\rho^{D-2}}\, u\theta(u)\right]^2\rho^2\,d\Omega_{D-3}^2 \label{m2},
\end{equation}
following from \rf{AiSe} by a coordinate transformation
\cite{D'Eath,EG,bh1}.
 Near the shock front, this metric can be approximated as
\begin{equation}\label{lin}
    ds^2\approx dx_\mu^2 +
    \frac{2{E}}{\Omega_{D-3}\,\rho^{D-2}}\,u\theta(u)\left[(D-3)d\rho^2-\rho^2\,d\Omega_{D-3}^2\right].
\end{equation}

Since, as previously mentioned, the center portions of the CTS can
be deformed away from the shock planes\cite{EG}, the region where
we have to check validity of the semiclassical approximation is
the region where the two pieces cross the shocks and join. This is
the region $\rho\sim R_h$ and at $|u|\ll R_h$.  Here the deviation
from the Minkowski metric is small:
\begin{equation}\label{linear}
    g_{\mu\nu}=\eta_{\mu\nu}+h_{\mu\nu},
    \qquad\left|h_{\mu\nu}\right|\ll 1,
\end{equation}
and we are in the linearized gravity situation.

To describe fluctuations, we will impose the transverse-traceless
(TT) gauge, specified by conditions %\cite{MTW}
\begin{equation}\label{TT}
    h_{\mu0}=0,\quad h_{ab,b}=0,\quad h_{aa}=0\ ,
\end{equation}
where $a=1,\ldots,D-1$.
We will also take advantage of the fact that the transverse shock
profile is slowly varying compared to the shock width $\Delta z$
in the region of interest. We can thus work in the plane wave
approximation \cite{bh1}, neglecting transverse derivatives of the
metric:
\begin{equation}
    \label{nonzero}
    h_{ab}\approx h_{ab}(u,v).
\end{equation}

Our goal is to estimate fluctuations of the shock wave amplitude
at $r\sim R_h$. The $h_{ij}$ corresponding to \rf{lin} can be
written as
\begin{equation}
    \label{pwa}
    h^{cl}_{ij} = \begin{pmatrix}
    (D-3) C u\theta(u) && && && \\
     && -Cu\theta(u) && && \\
     && && \ddots && \\
     && && && -Cu\theta(u)
     \end{pmatrix}.
\end{equation}
In the plane wave approximation the difference between polar and
Cartesian coordinates disappears. We can also neglect the
transverse dependence of $C$, so that it becomes a constant $\sim
R_h^{-1}$. After these simplifications, the classical field
\rf{pwa} becomes precisely of the form \rf{TT}, \rf{nonzero}.

We will estimate the strength of fluctuations of the gravitational
field compared to the semiclassical background in the region where
the CTS intersects the shocks. These fluctuations are
approximately controlled by the linearized Einstein-Hilbert
action, which in the TT gauge takes the form
\begin{equation}
    \label{act}
    S=\frac 18\int d^{D}x\, h_{ij,\ga} h_{ij}{}^{,\ga}.
\end{equation}
In the plane wave approximation this becomes
\begin{equation}
    \label{actpw}
    S=\frac A8\int dt\,dz \left[(\partial_t h_{ij})^2-(\partial_z h_{ij})^2\right],
\end{equation}
where $A$ is the transverse area of the  planar field
configuration being considered, here $A\sim R_h^{D-2}$.

Up to a constant factor this is the action of a 2-dimensional
massless scalar field. An estimate of the size of the quantum fluctuations follows straightforwardly from this:
\begin{equation}\label{}
    (\gd h_{ij})^2\sim A^{-1}\int_{k\sim \gd z^{-1} }\frac {dk}k\sim
    A^{-1}
\end{equation}

These fluctuations should  be much smaller than the typical size of
$h_{ij}^{cl}$ at a distance $\sim \gd z$ from the shock front,
which is $\sim C\gd z$ (see \rf{pwa}). This condition becomes most
restrictive when applied at the smallest scale $\gd z\sim \Delta z$
existing in the classical solution (shock wave width). Thus we get
the final condition for the smallness of quantum fluctuations:
\begin{equation}\label{lim2}
    A^{-1/2}\ll C\Delta z \quad\Leftrightarrow \quad \Delta z\gg
    R_h^{2-\frac{D}2},
\end{equation}
which is also compatible with \rf{limits}.

\section{Conclusions}

In this note we have argued that it is possible to carry out the
analysis of BH production in transplanckian elementary particle
collisions in a semiclassical approximation, taking into account quantum spreading of the wavepacket for a particle. The
crucial difference with \cite{R} (which led to the opposite
conclusion) was that instead of using the minimally allowed
uncertainty $E^{-1}$ as the particle wavepacket size $\Delta z$, we
noticed that it is unnecessary to insist on such a choice. and
kept $\Delta z$ as a free parameter. It was found that both criteria of
semiclassicality -- low curvatures and small quantum fluctuations -- can
be reconciled with the classical gravity analysis of BH formation,
provided that $\Delta z$ is in the range (see
\rf{limits},\rf{lim1},\rf{lim2})
\begin{equation}\label{}
    \max(R_h^{-1},R_h^{2-\frac D 2})\ll \Delta z \ll R_h.
\end{equation}
Note that, in practice,  to produce a black hole with $R_h\gg 1$, all this requires is a shock width larger than the  Planck size.  It is nonetheless interesting that a careful treatment of black hole creation requires quantum wavepackets, or classical continuous matter distributions.  This  analysis then puts the geometric cross section estimate on a more
solid ground.

\section*{Acknowledgements}
We would like to thank T. Banks, D. Eardley, and G. Horowitz for
important conversations. The work of SBG was supported in part by
the Department of Energy under Contract DE-FG02-91ER40618, and
part of it was carried out during the Workshop on QCD and String
Theory at the Kavli Institute for Theoretical Physics, whose
support is gratefully acknowledged. The work of VSR was supported
by Stichting FOM.

%%%%%%%%%%%%%%%%%%%%%%%%%%%%%%%%%%%%%%%%%%%%%%%%%%%%

\end{document}